\documentclass[twocolumn,aps,showpacs,preprintnumbers,amsmath,amssymb]{revtex4}

\usepackage{graphicx}

\begin{document}

\title{
{\large {\bf What one can learn from experiments about the elusive
transition state?}}}

\author{  {\bf
{Iksoo Chang$^{1,2}$, Marek Cieplak$^{1,3}$
Jayanth R. Banavar$^{1}$, and Amos Maritan$^{4,5}$\\
} }}

\affiliation{
$^1$ Department of Physics,
104 Davey Laboratory, The Pennsylvania State University, University
Park, Pennsylvania 16802\\
$^2$ National Research Laboratory for Computational Proteomics
and Biophysics, Department of Physics, Pusan National University,
Pusan, 609-735, Korea\\
$^3$Institute of Physics, Polish Academy of Sciences, 02-668
Warsaw, Poland\\
$^4$ INFM and Dipatrimento di Fisica `G. Galilei', Universit\'a
di Padova, 35131 Padova, Italy\\
$^5$ The Abdus Salam International Center for Theoretical Physics (ICTP),
Italy\\
}

\date{\today}




\begin{abstract}
{\noindent \bf Abstract}\\
We present the results of an exact analysis of a model energy landscape
of a protein to clarify the
notion of the transition state and the physical meaning
of the $\phi$ values determined in protein
engineering experiments. We benchmark our findings to various
theoretical approaches proposed in the literature for the identification
and characterization of the transition state.
\end{abstract}

\maketitle

\vskip 40 pt
\noindent {\bf Keywords:} protein folding, transition state, protein engineering
\vskip 40 pt



Small globular proteins are known to fold
rapidly and reversibly under physiological conditions 
(Anfinsen 1973)
This process is highly cooperative in nature
and is driven by hydrophobicity. It involves
expulsion of the solvent  from the interior of the protein's
folded state. The resulting  native state 
structure has a hydrophobic core which is stabilized by
hydrogen bonds and disulphide bridges. In the simplest case,
folding is an all-or-nothing phenomenon 
in that each individual protein molecule in a solution is either in a
folded (N, for native) or denatured (D) state and not in between.
This scenario is called a two-state picture
(Eyring and Stern 1939; Fersht 1998, Jackson and Fersht 1991; Otzen et. al.
1994, Itzhaki et al. 1995; Baldwin and Rose 1999a)
if it corresponds
to kinetics that is governed predominantly by a single exponential. 
The two-state picture is  anchored in the classic Eyring theory
(Eyring and Stern 1939)
of chemical reactions which envisions folding as
proceeding along a reaction coordinate so that the free energy
changes through three main stages 
(Fersht 1998, Baldwin and Rose 1999b):
D, $\ddagger$ -- the transition state, and N.
The transition state corresponds to the highest
free energy barrier and provides a bottleneck for the conversion
to the native state.\\

The phenomenological two-state picture raises many questions when
one considers the molecular structure of a protein.
For instance, there is a huge
number of conformations that the protein may adopt --
which of these ought to be classified  as $\ddagger$, or D? Do the other
conformations matter? What is the meaning of the reaction coordinate?
The transition state must be short lived and be barely populated so 
how can one find it experimentally or elucidate it theoretically?\\

One way to deal with the multiplicity of the microscopic conformations
is to view the folding phenomenon as being akin to a first order
phase transition (albeit in a finite system) with its kinetic 
mechanism being similar to nucleation
(Abkevich et al. 1994; Fersht 1997).
The notion of  
the transition state morphs then into that of a  folding nucleus
which acts as a critically sized droplet of the folded phase.
The criticality condition means that the droplet may either
shrink (which leads to unfolding) or expand (which leads to folding)
with equal probability, i.e. the droplet is on the edge between the
folded and unfolded basins of attraction. 
The nucleation interpretation
immediately suggests that there could be many different `droplets' that
form an ensemble of the transition states 
(Pande and Rokhsar 1999a; Pande et al. 1998; Pande and Rokhsar 1999b)
Is this suggestion valid?\\

The established experimental way to probe the transition state
or states is through the techniques of 
protein engineering
(Oxender et al. 1987; Robson and Garnier 1988; Cleland and Craik 1996;
Carmichael Wallace 1999; Fersht 1998; Matouschek et al. 1989; Matouschek
at el. 1990; Jackson and Fersht 1991; Otzen et al. 1994; 
Itzhaki et al. 1995).
The basic idea entails
the substitution of amino acids in different positions
of a protein with other amino acids
and monitoring the resulting changes in the stability of the 
native state and the kinetics of folding or 
unfolding.
The effects of these substitutions are characterized
by means of a set of the
folding $\phi$ values ($\phi _f$) which are measures
of the changes in the kinetic rates normalized by corresponding changes
in the protein stability.  In simple situations, the $\phi _f$'s
take the values between zero and one.
A value that is close to one suggests a nearly native-like
structure of the site of substitution in the transition state.
So new questions emerge -- for instance, how may one identify
conformations which are compatible with the measured $\phi$ values?
Furthermore, how may one interpret non-classical $\phi$ values which
are negative or bigger than 1? \\

The list of such basic and unsolved questions is long and so is
the list of different answers that have been offered in the literature.
This situation calls for considering a simple model that displays
two-state physics  and is amenable to exact solution, through which 
one may resolve the key issues and elucidate the underlying concepts.
In this paper, we analyze
a model that encapsulates many of the essential features of 
protein folding kinetics with a non-trivial 
free energy landscape. This model is a variant of a 
system considered by Munoz, Eaton and their collaborators
(Munoz et al. 1997; Munoz et al. 1998; Munoz and Eaton 1999).
It is Go-like (Abe and Go 1981)  
and it embodies the topology
of the $\beta$-hairpin. It was introduced in the context of experimental
studies of a corresponding fragment in the protein G (Munoz et al. 1997).
Munoz et al. (1997) have considered a peptide of 16 residues with one
tryptophan (W43) to investigate the kinetics of $\beta$-hairpin formation
in a laser-induced temperature jump experiment.  Measurements of the
tryptophan fluorescence have indicated that the relaxation to equilibrium
is governed by a single exponential and corresponds to a single free
energy barrier. The time constant is about 6 $\mu s$ which is about 30
times longer than that found in comparable $\alpha$-helices.
Its equilibrium properties have been further explored theoretically
by Flammini et al. 2002 and Bruscolini and Pelizzola 2002.
We consider a shorter, 12 amino
acid version of the original model, reformulate it
in terms of  Ising spins, which can take on one of two values,
corresponding to the immediate vicinity of the protein being native-like 
or not, and endow it with single spin flip kinetics -- Munoz and Eaton
had considered diffusional kinetics instead.
The kinetics are formulated in terms of a Master equation that deals
with probabilities and not specific trajectories.
We then go on to use the results of 
our exact solution of the model to understand the nature of
the transition state and the significance of $\phi$-values.
\\

\noindent
{\bf Results}\\

\noindent
{\it The model}\\

The native state of the system we study is illustrated in Figure 1.
The system can be described in terms of effective free energy levels
which take into account their underlying microscopic
degeneracies through an effective entropy term. The free energy
levels are defined in terms of 11 peptide bonds which are either
placed in the native fashion or not. 
The native placement corresponds to the Ramchandran
$\phi$-$\psi$ angles taking on their native state values.
This binary character of the
bond placement allows for an Ising-like modeling and we adopt 
spin variables $S_n$ which take values 1 or 0 correspondingly.
The free energies per mole can be written as
\begin{eqnarray}
G \;  =\; -J\sum_{l<m} \Delta_{lm} \prod _{n=l}^m \; S_n \;+\;
T \Delta S_{conf} \sum _{n=1} S_n  \;\;\;\;\;\; & & \nonumber \\
    =\; -J ( S_5 S_6 S_7  \;+\;S_4 S_5 S_6 S_7 S_8 & &  \;\;\; \\
    + S_2 S_3 S_4 S_5 S_6 S_7 S_8 S_9 S_{10} 
    + S_1 S_2 S_3 S_4 S_5 S_6 S_7 S_8 S_9) & & \nonumber \\
   -2J(S_1 S_2 S_3 S_4 S_5 S_6 S_7 S_8 S_9 S_{10} S_{11}
    + S_3 S_4 S_5 S_6 S_7S_8 S_9)  & & \nonumber \\ 
  +T \Delta S (
    + S_1 +S_2 +S_3 +S_4 +S_5 +S_6 +S_7 +S_8 + S_9 + S_{10} +S_{11}) & &
   \nonumber
\end{eqnarray}
A non-zero value of the
product $S_l S_{l+1} ... S_m$ implies that all peptide bonds
between $l$ and $m$ are set in the native fashion which allows 
for the establishment of native interactions in the cluster between
the bonds $l$ and $m$. These interactions are either hydrophobic
or due to establishment of the hydrogen bonds or
both. For simplicity, we assume that the strength of the interactions, 
$J$, are the same in both cases and equal to 
1000 K whereas the conformational
entropy per spin, $\Delta S_{conf}$, is taken to be 2.14 $R$, where
$R$ is the gas constant -- in the equation above, $T$ denotes 
the temperature. 
Therefore, the stability of the $\beta$-hairpin system is controlled by a
competition between the gain in the energy of the established contacts and
the loss of conformational entropy on setting the conformational angles to
their native values.
We choose
$\Delta _{lm}$ to be  2 for $(l,m)$=(1,11) and (3,9), 1 for $(l,m)$=(2,10),
(4,8), (5,7) and (1,9), and 0 otherwise. 
Note that the placement of the contacts breakes the symmetry
between the upper and lower branches of the hairpin.
There are several reasons why we consider the simpler 12-residue system.
First, the number of conformations is significantly smaller which
facilitates the computational study.  Second, the model is simplified by
choosing just one interaction parameter. Third, we remove an unnecessary
complication of the original model which is related to the fact that two
residues at each of the terminal ends of the hairpin are not stabilized by
the interactions present in the 16-residue system. In fact, the haipin
conformation is not a free energy minimum for either the original or the
simplified couplings.
Let the free energy in conformation $i$  be denoted by
$G_i$. The equilibrium probability to occupy this conformation,
$P_i$ is then given by
\begin{equation}
P^{eq}_i \;=\; \frac{e^{-G_i/RT}}{ \sum _i  e^{-G_i /RT}} \;\;.
\end{equation}
\\

\noindent
{\it Kinetics: relaxation, folding and unfolding}\\

The relaxational spin-flip kinetics can be described in terms of the
Master equation (see, e.g., Cieplak et al. 1998; Ozkan et al. 2002)
for the time dependent vector of probabilities $\vec{P}$ with 
components $P_i$. By convention, we take $i=1$ to correspond to the
native state. The Master equation reads
\begin{equation}
\label{master}
\frac{d}{dt}\vec{P} \;=\; -M \vec{P} \;\;,
\end{equation}
with $M_{ij}$ for the flip from state $j$ to $i$ equal to
-$\frac{1}{\tau _0}$ if $G_i < G_j$ and $\frac{1}{\tau _0}
exp-(G_i - G_j)/RT$ otherwise. $\frac{1}{\tau _0}$ is the attempt rate
which may generally depend on $T$.
The diagonal elements are set so that
the sum of the terms in each column is zero. This choice of the
matrix is consistent with the detailed balance condition and the
Arrhenius form of the low-$T$ relaxation processes.
The time evolution of
$\vec{P}$ can be obtained through an iterative use of the
equation $\vec{P}(t+\delta t) = (1 - \delta t M)\vec{P}(t)$, where
$\delta t$ denotes an infinitesimal time increment.
An alternative way to follow the kinetics is by decomposing
$\vec{P}$ into the right-handed eigenvectors and by endowing them
with an exponential time dependence of the form $exp(-\lambda _{\alpha}t)$,
where $\lambda _{\alpha}$ are the eigenvalues of the $M$ matrix.
One eigenvalue is always zero - it corresponds to the system staying
in equilibrium. The smallest non-zero eigenvalue,
denoted as $k$, is the slowest relaxation rate. 
The inverse of $k$ yields the longest relaxation time.
Other eigenvalues correspond to faster processes.
The two-state behavior is obtained when there is a substantial
separation between the slowest and other rates. Such is indeed
the case here since our choice of the parameters yields the second
longest relaxation time at 300 K to be of order 6\% of the longest one.
Folding conditions are generated when one disallows all transitions
that lead out of the native state - the first column of the $M$ matrix
is set equal to zero and the native state acts as the
probability sink. The resulting matrix will be denoted by $M_f$.
On the other hand, the unfolding conditions are
generated by making the completely unfolded state
a probability sink
and the corresponding column is set equal to zero to obtain the
matrix $M_u$. In these
cases, the smallest non-zero eigenvalue corresponds to the
slowest folding and unfolding rates, denoted
as $k_f$ and $k_u$ respectively.\\


The free energy levels in our model depend on temperature. However, in
order to identify kinetic barriers, it is useful to freeze the levels at
their 300 K values and to introduce another fictitious temperature, $T'$,
that can be varied at will. In particular it can be set to zero to
identify characteristic times that diverge in this limit. We find that the
eigenvalues (the inverse relaxation times) either tend to zero as $T'$
approaches zero (there are four such eigenvalues with eigenvectors
corresponding to the 4 local minima) or they tend to integer values of 1,
2 or 3. The other eigenvalues correspond to downhill motion in the free
energy landscape and are determined by the its local topology. As $T'$
increases there are Arrhenius-like corrections to the $T'$=0 limit and
considerable mixing of the levels.\\

Before we continue with the discussion of our model, we note
that a strictly two-level system would be described by the
following $2\times 2$ matrices
{
\begin{equation}
M \tau _0 =
\left [
\begin{array}{cccc}
k_u &  -k_f \vspace*{0.5cm} \\
-k_u &  k_f  \vspace*{0.5cm} \\
\end{array}
\right ] \;\;
M_f \tau _0 =
\left [
\begin{array}{cccc}
0 &  -k_f \vspace*{0.5cm} \\
0 &  k_f  \vspace*{0.5cm} \\
\end{array}
\right ] \;\;
M_u \tau _0 =
\left [
\begin{array}{cccc}
k_u &  0 \vspace*{0.5cm} \\
-k_u &  0  \vspace*{0.5cm} \\
\end{array}
\right ] \;.
\end{equation}   }

The transition state is implicit and  $k_f$ and $k_u$
satisfy
\begin{equation}
\label{kf}
k_{f,u}\;=\; \frac{1}{\tau _0}
exp(-\frac{\Delta G_{f,u}^{\ddagger}}{RT}) \;\;,
\end{equation}
where
$\Delta G_f^{\ddagger} = G_{\ddagger} - G_D$ and
$\Delta G_u^{\ddagger} = G_{\ddagger} - G_N$.
Each of the $M$ matrices has one zero eigenvalue and the other
eigenvalues are $k=k_f+k_u$, $k_f$, and $k_u$ for the relaxation,
folding, and unfolding situations respectively which agrees
with the standard expectation (Fersht 1998). 
The eigenvector
corresponding to the non-zero eigenvalue is, in each case, equal to
$
\left(
\begin{array}{c} 1 \\-1 \end{array}
 \right)
\;$. Thus  the observation of Ozkan et al. (2002) 
that the populations corresponding to the relaxation eigenvector of the
lowest non-zero eigenvalue are `rigorously what should be called 
transition state conformations' is not valid.
In the two-level system, the eigenvector contains both the
$D$ and the $N$ conformations, albeit with opposite sign.
One can show exactly that, quite generally, the sum
of the components of the eigenvector is zero and corresponds to a draining
out of the probability of occupancies of conformations with a given sign
in the eigenvector accompanied by an associated increase in the probabilities 
of the remaining conformations. \\

In the 12 amino acid model, there are
2048 possible conformations. The native state
corresponds to all spins being equal to one and
to the establishment of all 8 contacts.
The fully unfolded state corresponds to all
spins being zero and no contacts. Of these 2048 conformations,
67 of them
have the property that non-zero values of the spins are contiguous,
i.e. form a single sequence of ones. Indeed, the 67th conformation
is the unfolded state in which all the spins have zero values.
In the so called
single sequence approximation (Munoz et al. 1998) 
one restricts the conformation space to
just these 67 states. 
Figure 2 shows that the single sequence approximation gives a
fairly accurate picture of the thermodynamic quantities.
For the parameters chosen, the folding temperature, $T_f$, 
is around
300 K, both exactly and in the single sequence approximation,
and this is the temperature at which we focus our
further studies. At $T_f$, the probability to occupy the
native state is around 1/2 and the specific heat and fluctuations
in the fraction of the established native contacts, $Q$, shows a
maximum.\\

The 67 states of the single sequence approximation
are shown in Figure 3 in a form of a triangle.
The single circle at the bottom represents the
unfolded state (state 67).
The  bottom row of circles represents
states with one non-zero spin. The second row represents states
with two contiguous non-zero spins, the third - with three, and so on.
The top circle represents the single
native state (state 1). The kinetic moves from the unfolded state
(the bottom state) can
connect to any of the single spin states (last but one row)
and vice versa. In all other cases,
the allowed kinetic moves are only along the diagonal directions
on the triangle, as shown by the dotted lines around the 58'th state.
There are at most four possible moves because the single
sequence condition allows for changes occurring exclusively at the
interface(s) between the spin ones and the spin zeros.\\

\noindent
{\it The free energy landscape: the transition state}\\

The free energy landscape, of course,  depends on the parameters of
the model, especially the $T$.
The top panel of Figure 3 illustrates the principal features of the
free energy landscape at
$T$=300 K. There are three local minima, denoted by the larger
double circles (states 24, 34, and 43), and two local maxima, denoted
by the smaller double circles (states 6 and 46). One can
ask in which direction, preferentially towards the native or
towards the unfolded state, does the flow of the
probability occur were only one state  occupied initially.
This propensity can be straightforwardly determined by making both
the first and last columns of the $M$ matrix equal to zero
and by studying
the time evolution of the probabilities. In this way, both
the native and unfolded states act as probability sinks.
We find that all states at the top of the triangle have a strong
preference to flow toward state 1 (i.e. to reach $P_1 \approx 1$
and $P_{67} \approx 0$) whereas all bottom states
produce a flow to state 67. There are six states ``on the edge"
(5, 16, 25, 33, 40, and 39, shown connected by a line in the figure)
which show nearly equal propensities for both directions and
they separate the two regions of behavior.
Two of the edge states, 25 and 33, are ``confused" the most and they
also have the lowest free energy among the six. Strikingly,
none of the edge states
is a maximum.\\

The bottom panel of Figure 3 illustrates the best paths that
allow for the optimal pathway between states 1 and 67, in either direction.
They correspond to the states marked by stars within the circles.
There are 48 such paths
because on several horizontal lines
of the figure there are several
states to choose from. These choices are equi-energetical
making the optimal path energetically unique. The corresponding
free energies and the numbers of the native contacts formed are
shown in the right of the bottom panel. 
The native state corresponds to the free energy of -938 K.
The highest barrier
to climb on the best trajectories is 1852 K and it corresponds
to populating two degenerate states: 25 and 33 which are the saddle point
states. There are
two native contacts that are established in these states:
between bonds (or spins)
5-7 and 4-8, i.e., $S_4=S_5=S_6=S_7=S_8$=1. In state 25, $S_3$
is non-zero whereas in state 33 it is $S_9$ which is non-zero.
Thus these two of the edge states are the {\em transition states}.
and are shown in the figure as  the black circles. 
The identification of states 25 and 33 as transition states comes also
as a result of studies of sensitivities of $k_f$ and $k_u$ to changes
in the free energy values of individual conformations and noting that
the influence
of such perturbations is largest in the transition states.
This large sensitivity is due to the fact that the transition states
act as bottlenecks for the folding and unfolding kinetics.
When one considers the full
2048-level description there are 11!=39 916 800 different directed
paths from (00000000000) to (11111111111). Among these, there are
432 optimal trajectories which include the 48 identified in the
smaller subset of states -- the transition states are the same.
In the full set, there are four other conformations having 
the same energy as the transition states, 
e.g. (10011111000), but the optimal directed paths do not encounter
these conformations. \\

The reaction coordinate for the folding transition consists of a list
of conformations that are travelled on a directed optimal trajectory.
The free energy plotted against this
reaction coordinate is shown in Figure 4 (bottom panel). 
It indicates states 25 and 33 as the transition states.
This plot is quite distinct from the free energy, $G(Q)$,
calculated as a function of the fraction of the native contacts.
As seen in the top panel of Figure 4, $G(Q)$ has a maximum
for $Q=\frac{1}{4}$ which corresponds to seven states,
but only two of them, 25 and 33, 
are actually the transition states
as obtained through the studies of the kinetic connectivities.
We note that the choice of $Q$ as a reaction
coordinate has been made, for instance, by Munoz et al. (1998),
Clementi, Nymeyer, and Onuchic (2000) 
and Shea, Onuchic, and Brooks (2000).
They considered Go and non-Go off-lattice models
with strong dihedral angle terms in the potential energy. These models
exhibit a double minima structure in the free energy when plotted
against energy or the fraction of the native contacts that are
established during the time evolution of molecular dynamics simulations.
They assumed that states contributing to
the maximum separating the two minima (i.e. those which are
``half way" in terms of the number of contacts) form the 
transition state ensemble. \\

It should be noted that, in our model,
the proper reaction coordinate emerges
naturally when arranging the states according to their magnetization,
i.e. the net spin value, and not $Q$. The kinetic connectivities
relate neighboring values of the magnetizations which translates into
complicated connectivities between states of a given value of $Q$.
This is illustrated in Figure 5 which shows, in particular, that
the transitions link both same and distant values of $Q$ indicating
no simple relation to the transition coordinate.\\

It is not straightforward to discern the transition state from the
eigenvectors.
The smallest non-zero relaxation
eigenvalue corresponds to an eignevector with a mixture
of positive and negative components, which add to zero.
The transition state conformations
come with a weight of the same sign as the native conformation
and with an opposite sign to that
of the unfolded conformation.  The eigenvectors corresponding
to the smallest non-zero eigenvalue for folding and unfolding
have just one component of one sign (native and unfolded respectively)
with the transition state being undistinguished and ranked around 40th
among the remaining 66 conformations. \\

\noindent
{\it Time evolution of the probabilities}\\

Our framework provides a straightforward mechanism for monitoring
the temporal evolution of the probabilities of the protein to be in a given
conformation and average values of physical quantities
in terms of linear combinations
of the eigenvectors of the $M$ matrix. The smallest non-zero eigenvalues
describe the long time behavior. The combined effect of all eigenvectors,
at any time, can be assessed from the full time evolution of
$\vec{P}$. Figure 6 shows the evolution of $P_1$ and $P_{67}$ in the
67-level system under the conditions of folding, unfolding and
relaxation. 
The plots for folding and unfolding are not symmetric:
the occupation of the unfolded state disappears much more rapidly
on folding than of the native state on unfolding. This is because
there are many more ways to exit from state 67 compared to just two ways
to exit from state 1 leading to much smaller contributions from state 
1 to the eigenvectors corresponding to large eigenvalues (or short times). \\ 
Figure 7 shows a similar plot for the transition state 33 and 40, a
neighboring conformation.
Note the disparity in the scales of the $y$-axis in Figures 6 and 7.
In general,  there is nothing in the time evolution
of the probability of occupancy of  the transition state that
would distinguish it from any other states 
(with the exception of 1 and 67).
The maximum values reached by $P_{33}$ and $P_{25}$ are on the lowest side
when compared to other states. Thus the likelihod that they would be spotted
in a computer simulation is very low.\\

\noindent
{\it The chevron plots}\\

We now focus on the long time evolution, as determined from the
smallest eigenvalues. The experimentally measured kinetic rates 
are usually represented as the so called chevron plots 
(Fersht 1998; Chan and Dill 1998)
in which the logarithms of the rates are plotted against the concentration
of a denaturant. The couplings used in our model are meant to
correspond to  physiological conditions. 
In order to mimic  the effects of a denaturant, 
we adjust the coupling $J$ in a linear fashion so that
$J(x)= (1-x)J $, i.e. $x$ is assumed to be equal to the
fractional change in $J$ compared to its $x$=0 value
(see Figure 4). Figure 8 shows that
the resulting plots of the logarithms of the rates vs $x$ are
chevron-like with some curvature in the branches. The relaxation
curve agrees approximately with the condition $k=k_f+k_u$ which arises in the
two-state picture. Furthermore, it is seen that
the 67-level data points are well described by a  
system reduced  just to four levels : 1, 25, 33, and 67. 
Considering the
full set of 2048 states affects the folding branch very little
but it shifts the unfolding branch (and thus also the
relaxation curve): more states allow for a faster unfolding in analogy
to the asymmetry discussed in the context of Figure 6, because
there are more states to go to from the native state.
Nevertheless there is no qualitative distinction between the
chevron plots for the 67- and 2048-level systems other than the
location of the $x$ value at which the folding and unfolding curves
intersect. \\

The linear adjustment in the $J$ coupling appears to be 
a plausible model
to study the effective influence of $x$. Another simple model that can
be considered is to introduce the
free energy adjustments that are coupled to $Q$ and are thus
cooperative in nature. One way to do it is
to take $G_i(x) = G_i + |G_i|Qx$. The corresponding
chevron plot is shown in Figure 9. The 
folding and unfolding branches are seen to be straighter
but the overall character of the $x$-dependence is qualititatively 
similar to that shown in Figure 8.\\

{\it The $x$-dependence of $\beta ^{\ddagger}$}\\

We now consider the $x$-dependence of the slopes in the
chevron plots. We define
\begin{equation}
m_{f,u}\;=\; \mp \frac{\partial ln(k_{f,u})}{\partial x} \;\;
\end{equation}
and similarly
$m\;=\;\frac{\partial ln( K)}{\partial x} \;$, where $K$
is the equilibrium constant which in the two state model is given by
\begin{equation}
\frac{1}{K}\;=\;\frac{k_f}{k_u}\;=\;\frac{P^{eq}_N}{P^{eq}_D}
\end{equation}
The two-state picture holds if
$m\;=\; m_f + m_u \;$. At $T_f$, $K$ should be 1 and this is nearly
the case if we count not only state 67 but also all of the last but
one row states of the triangle of Figure 3 as belonging to the
coarse-grained denatured state.
Another quantity of interest 
is the parameter $\beta ^{\ddagger}$
defined as
\begin{equation}
\label{eqba}
\beta ^{\ddagger}\;=\frac{|m_f|}{|m_f|\;+|\;m_u|} \;\;.
\end{equation}
Let us postulate 
a linear effect of the denaturant's concentration on the
free energies so that
$G_N(x)\; =\; G_N + y_N x \;$ and 
$G_{\ddagger}(x) \; = \; G_{\ddagger} + y_{\ddagger} x  \;$, where
$G_i$ ($i=N, \ddagger , D$) on the right hand sides of the equations
denote the $x$=0 values of the free energy; the denatured state
is expected to be unaffected by $x$.
In this case,
$\beta ^{\ddagger} \;=\; y_{\ddagger}/y_N \;$,
i.e. $\beta ^{\ddagger}$ does not depend on $x$. 
This expression for $\beta ^{\ddagger}$ indicates that this
quantity measures the amount of the native state-like structure
contained in the transition state which in turn suggests the
common interpretation that it
is related to the amount of the buried surface area.
There are proteins,
however, in which $\beta ^{\ddagger}$ shows a linear variation
with $x$.
A varying $\beta ^{\ddagger}$ would mean then
that either the free energy of the transition state varies with $x$
in a way unrelated to the free energy changes in the native state
(e.g. because of the
presence of non-native contacts in the transition state)
or that the identity of the transition state
varies with $x$. In the latter case, the adjustments of the
free energy landscape can be captured by a `movie' 
(Oliveberg et al. 1995; Ternstroem et al. 1999; Oliveberg 2001).
In our model, the transition
state remains the same, i.e. it does not ``move", when $x$ changes
between -0.25 and 0.25, as shown in  Figure 4, and yet
$\beta ^{\ddagger}$ varies. The bottom panel of Figure 10
shows that the dependence is nearly linear. The slopes in the
67- and 2048-level systems are almost the same. It is only in the
limit of four states that $\beta ^{\ddagger}$ is constant, 
and equal to $\frac{1}{4}$.
If all states are included, the chevron branches
acquire curvature (see Figure 8) and $\beta ^{\ddagger}$ is merely
a measure of the curvature generated by the presence of states
which are not present in the two-state picture.\\

\noindent
{\it The $\phi$-values}\\

In order to determine the analog of the $\phi$-values
in our model, at $x$=0, we consider a small
local adjustment in $J$ at the  location of a given
amino acid. The adjustment is taken to be of order
5\%. The $\phi$-values are practically independent of the
magnitude of adjustment between 1 and 5\%.
There are 12 possible locations which are either
at a joint between two bonds (two spins) or at the
end points of the system.
Note that various amino acid locations correspond to
different numbers of bonds that are affected.
For instance, the ninth amino acid belongs to bonds $S_8$ and $S_9$
(see Figure 1) which are coupled to  four interactions
that are affected as a result of a `mutation' on this site.
Each  adjustment affects the folding and unfolding rates
by $\delta k_f$ and $\delta k_u$ respectively
which allows one to calculate 
\begin{equation}
\label{eq3}
\phi _f = \frac{\delta ln (k_f)}{\delta ln (k_f/k_u)}\;=\;
\frac{\delta k_f}{k_f}\;/\;(\frac{\delta k_f}{k_f}
-\frac{\delta k_u}{k_u})
\end{equation}
and
\begin{equation}
\label{eq33}
\phi _u = \;-\frac{\delta ln (k_u)}{\delta ln (k_f/k_u)}\;\;
\end{equation}
for a mutation at any of the 12 amino acid sites. Note that the
folding and unfolding $\phi$-values satisfy the condition
$\phi _f + \phi _u \;=\;1 $. \\

The two state picture interprets
the $\phi$ values in terms  of changes in the
Gibbs free energy of the folded state and the transition state
brought about  by the mutation. Specifically, using eq. (5),
\begin{equation}
\label{eq4}
\phi _f \;=\; \frac{\delta \Delta G_f^{\ddagger}}{\delta \Delta G}\;\;
\end{equation}
and
\begin{equation}
\label{eq5}
\phi _u \;=\; -\frac{\delta \Delta G_u^{\ddagger}}{\delta \Delta G}\;\;,
\end{equation}
where the symbol $\delta$ indicates a change in, say,
$\Delta G= G_N - G_D$ relative
to the respective wild type value. 
The two state picture is obtained when one restricts the 
conformation space to just four levels: 1, 25, 33, and 67.
The set of the corresponding
$\phi _f$ values is shown in Figure 10 as asterisks and marked
as 4-state. They are equal to 1 at sites 5, 6, 7, and 8
(between bonds $S_4$ and $S_8$); equal to $\frac{1}{3}$ at site 4; to
$\frac{1}{4}$ at sites 9; and to 0 at the remaining end sites.
This pattern is consistent with the structure of states 25 and 33.
When we consider 67 levels, the sites near the turn still have
high $\phi _f$-values but they become reduced to about 0.8. At
the same time, the values near the end points are enhanced and only
the very end points continue to have strictly vanishing  $\phi$ values.
The pattern of the $\phi _f$ values gets a small shift when the
full set of 2048 states is considered. It should be noted that
the $\phi$ values depend on $T$ and on other modifications in
the free energy landscape such as a  lowering of one of the free
energy minima.\\



\noindent
{\bf Discussion}\\

There are a number of approaches to interpret the transition state
in fast folding proteins in which no intermediates are involved. 
We have already discussed some of the concepts and results.
These are: 1) the reaction coordinate is neither $Q$ nor another
macroscopic observable but a list of conformations
travelled on the optimal trajectories,
2) the transition state/states can be
identified by enumerating possible trajectories, 3) the transition 
states are substates of the edge states
which are as likely to fold as to unfold, 4) transition states are
not easily determined by the eigenvector of the $M$ matrix
corresponding to the longest relaxation time, 5) an $x$-dependent
$\beta ^{\ddagger}$ does not indicate a moving transition state.
\\

The free energy landscape
of our model is not endowed with a funnel (Onuchic et al. 1995)
and yet it
provides for fast folding. 
Whether the landscape incorporates a funnel or not, one would expect
that the transitions states are akin to saddle 
points with very low occupational probabilities. 
Such states ought to be hard to spot through simulations.\\

It should be pointed out that studies of the so called disconnectivity
graphs for the polyalanines 
(Dobson et al. 1998; Becker and Karplus 1997; Levy et al. 2001)
also do not yield a funnel-like landscape and suggest instead that
the conformational space should be visualized as a
broad basin with several pronounced minima at its bottom. The
disconnectivity graphs constructed by Wales et al. 
(Wales et al. 2000)   
for various protein-like systems are endowed with many ``transition states".
These, however, are defined as saddle point conformations
separating two arbitrary local energy minima.
One of these saddle points should correspond
to the transition state of Eyring but all others are not
expected to be relevant kinetically.\\

\noindent
{\it The issue of multiplicity of folding nuclei}\\

A multiplicity of distinct transition states or critical droplets 
is also implied by the nucleation-condensation picture of folding
(Abkevich et al. 1994; Fersht 1997)
and the neo-classical approach of
Pande, Rokhsar and their collaborators
(Pande and Rokhsar 1999a; 1999b; Pande et al. 1998)
In practice, the droplets were identified as the edge
conformations such that time evolution
leads to folding and unfolding with equal probabilities.
In lattice models, 
these probabilities are calculated by determining the fate of enumerated
short Monte Carlo trajectories that originate from the conformations.
Our calculations show that only the lowest free energy edge states
are transition states. 
Pande and Rokhsar have also studied off-lattice models through
all-atom molecular dynamics simulations in unfolding trajectories.
In particular, they considered the $\beta$-hairpin system of protein G
(Pande and Rokhsar 1999b) 
(related studies were done in Dokholyan et al. 2000 and Ding et al. 2002) 
and identified four characteristic stages - or clusters
of conformations - denoted consecutively as F, H, S, and U. They 
identified conformations (regions of values of the radius of gyrations
and of $Q$) which correspond to the edge states separating F and H
and similarly the edge states separating S and U. Both are treated
as independent transition states without a comparison of their
free energies and without a determination of the edge region
between H and S.  The edge region between H and S may actually correspond
to the highest energy and if so it would correspond to the true
transition state provided the paths which go through the stages F-H-S-U 
are close to being optimal. The procedure of determining
`transition states' for pairs of certain stages may not be always correct
because the problem of the
optimal path is global in nature and 
partitioning it into sub-tasks may work only as an approximation.\\

We should also point out that their procedure identifies
the hydrophobic cluster (in our model, spins 1,2,3,9,10, and 11) as
folding first and the turn region as folding last. This does not agree
with either the original interpretation of the experiment
(Munoz et al. 1997,1998) or with the structure of the transition state
found in our model. It is interesting to note, however, that an
all-atom multicanonical Monte Carlo simulations with implicit
solvation effects performed by Dinner et al. (1999) suggests that the
folding does indeed start at the hydrophobic cluster.
Furthermore, the folding rate is found to be dominated by the time
scale of interconversion between compact conformations. Although the
experiment (Munoz et al, 1997, 1998) does not exclude this folding
scenario, the additional experiments and simulations may yield a more
complete understanding of the folding kinetics in the $\beta$-hairpin.
Our model is not meant to generate a realistic picture of the hairpin
but is meant to merely provide an illustration of the concepts.\\

\noindent
{\it Transition state through abrupt changes in the structure}\\

The picture of multiple folding nuclei
has been also advocated by Klimov and Thirumalai (2001) 
They also argued that these nuclei should contain non-native
contacts. Our analysis does not allow us to draw any conclusions about
the role of the non-native contacts because they are not
addressable in the present model. 
Their method of identification of the folding nucleus is
based on sudden changes in structure in the very last
stages of folding, i.e. when the time evolution ought to be entirely
governed by the eigenvector corresponding to the smallest folding
eigenvalue which has very little weight in the transition state.
Note that there are
no sudden changes in properly averaged time dependent
observables, as evidenced in our model by Figures 6, and 7.
In particular, the probabilities  to establish contacts are given
by curves which are smooth and monotonic.
Thus any abrupt features should be either due to the presence
of intermediates (i.e. be outside of the two-state picture) or
be due to insufficient averaging. If one trajectory shows an
abrupt structural change at one point, there must be other
trajectories which would have abruptness at other points
so that a many trajectory average is smooth.\\

A similar criticism applies to the molecular dynamics based identification
of the transition state 
(Li and Daggett 1994; Kazmirski et al. 2001).
The operational definition of the transition states is given
``as the ensemble of structures populated immediately prior to the
onset of a large structural change" during unfolding. 
Note that all sufficiently averaged
quantities should be smooth functions of time, as discussed above.
Thus any  method based merely on abrupt changes in the structure
probably cannot identify the transition state.
Furthermore, it should be noted that unfolding simulations typically
impose unfolding conditions through an
application of a high temperature (above 200 C) and
sometimes high pressure. Both of these circumstances are expected
to alter the free energy landscape significantly -- possibly beyond
any meaningful comparison with the experiment.\\

\noindent
{\it The reaction coordinate and eigenvectors}\\

We have already mentioned the attempts to link the transition
state to the eigenvalue analysis 
(Ozkan et al. 2001; Ozkan et al. 2002) 
of the Master equation.
They argue that the eigenvector corresponding to the lowest
eigenvalue of the relaxational $M$
matrix can be interpreted as providing 
a  reaction coordinate and a selection
of the transition state.
We find in our model that the relaxational eigenvector is a linear 
combination of essentially all 67 conformations and the true transition
state is the 12'th weakest weight state.\\ 

\noindent
{\it Selection of the transition state based on the $\phi$-values}\\

An entirely different way to determine the transition state
is generated by a computational exploration of the conformations
of a protein followed by an attempt to match them
with experimentally determined $\phi _f$-values. 
If the models are off-lattice then the procedure involves some
clustering of conformations.
Example of this
approach are in papers by Vendruscolo et al. (2001) 
and Paci et al. (2003) 
The assumed connection of a conformation with the $\phi _f$ 
values is through the degree of nativeness, $\kappa _i$, 
of the local structure. This degree is defined
by the number of established native contacts
that are linked to the mutated amino acid  divided by the 
maximal native number.
The calculated values of $\kappa _i$
are then compared to the experimental values $\phi _i$ which
are defined as $\phi _f$ at site $i$.
The transition state conformations are assumed to be those which
minimize the distance between $\kappa _i$ and $\phi _i$.
Paci et al. (2003)   
have found a dynamical way of generating the
best conformations of this sort by running a simulation which
punishes the departures from the experimental values of $\phi _i$.\\

It is easy to test this approach in the 67-level model.
We determine the $\kappa _i$ values and compare them to the $\phi _i$
obtained through the Master equation approach. We find
that there are seven conformations which have the smallest
and identical Euclidean distance of 0.636 from the kinetically
derived values. In addition to the two transitions states 25 and 35
these are states 4, 15 31, 32, and 34 defined as
(11111111000), (01111111000), (00011111111), (00011111110), and
(00011111000). 
These seven conformations form a V shape 
in the diagram of the states shown in Figure 3.
All of them have the $\kappa $ values given by
(0 0 0 $\frac{1}{3}$ 1 1 1 1 $\frac{1}{4}$ 0 0 0).
Our conclusion is that even though the $\kappa$-value based
method does succeed in finding the transition states it
also finds many other spurious  conformations.
We conclude that this approach is not fool-proof if it is not followed
by some additional selection of the states.
The need for additional criteria for the selection of transition
state conformations was highlighted by Vendruscolo et al.
(2001) who used the $\beta$ Tanford analysis for this purpose.\\


\noindent
{\it Non-classical $\phi$-values}\\

We now consider the issue of the non-classical, i.e. negative
or bigger than 1, values of $\phi$.
Ozkan, Bahar, and Dill (Ozkan et al. 2001;2002) 
argue that the folding pathways have a different character away
from the native state, where there is a multiplicity of
parallel ``routes downhill", and near the native state,
where folding is slow and serial-like. They postulate that the
transition state is located near the place where there is a
change in the network topology and it acts as a switch for the
flows of probability.
They consider a specific model which is assumed to have two main
channels for the flow and
suggest that mutations may destabilize one, say slow,
channel and direct more flow to another channel. This picture
allows them to argue that the $\phi$-values are measures
of the acceleration/deceleration of folding resulting from
the mutations.
Their model yields non-classical values of $\phi$.
\\

Consider a folding rate that is a sum of two independent, parallel  processes
(i.e. of the probability flow through two channels): $k_f\;=\;
k_{f1} + k_{f2}$ and similarly $k_u\;=\; k_{u1}+k_{u2}$. 
We assume that the single channel folding and unfolding rates
are described as in eq.(\ref{kf}) but with
the individual barrier heights $\Delta G^{\ddagger}_{fi}$
and $\Delta G^{\ddagger}_{u\gamma}$ ($\gamma$=1,2). Suppose that
a mutation shifts the native state free energy by $g$
so that
\begin{equation}
G_N\;=\; G_N^0 + g \;\;,
\end{equation}
where the superscript 0 indicates the wild type value.
We assume that the mutation does not affect the free energy of the
denatured state, $G_D = G_D^0$, whereas the individual transition
state free energies get shifted in proportion to $g$. Thus
\begin{equation}
G_{\gamma}^{\ddagger}\;=\; G_{\gamma}^{\ddagger 0} \; + \mu _{\gamma} g \;\;,
\end{equation}
where $\mu _{\gamma}$ are the coefficients of proportionality. 
Note that
eq. (\ref{eq3}) can be rewritten as $\phi _f \;=\; 
[1 -\frac{k_f\delta k_u}{k_u \delta k_f }]^{-1}$.
In the two channel case,
\begin{equation}
\frac{\delta k_u}{\delta k_f}\;=\;
\frac{(\mu _1 -1)k_{u1} \; + \; (\mu _2 -1) k_{u2}}
     {\mu _1 k_{f1} \; + \; \mu _2 k_{f2}} \;\;.
\end{equation}
Note that the coefficients $\mu _{\gamma}$ are expected to be less than one
and positive which means that $\frac{\delta k_u}{\delta k_f}$ is negative
and thus $\phi _f$ cannot exceed 1. 
A possibility for
non-classical values of $\phi$ would arise were the coefficients to
have opposite
signs. This could arise naturally 
when the transition state has non-native contacts,
as noted by  Li, Mirny and Shakhnovich (2000).  
In the case of the three state barnase, the non-native contacts have
been revealed through protein substitution studies
(Matouschek et al. 1992; Tissot et al. 1996; Dalby et al. 1998)
as arising in a long lasting intermediate state.
\\

\noindent
{\bf Conclusions}\\

Our benchmarking of various methods to determine the transition
state in the exactly solvable model indicates that the most
practical method entails using the experimental values of $\phi$
combined with kinetic simulations
to determine the set of conformations which are both the most
compatible with the $\phi$-values and are edge states.  A further
refinement would entail picking conformations with the lowest free energy 
from this predetermined set.
Such a refinement is probably less necessary for a large protein
with multiple constraints imposed by the $\phi$ values.
It is possible that for sufficiently large proteins compatibility
with the kinetic simulations may already select the correct state.

\vskip 40 pt
\noindent
{\bf Acknowledgements}\\

\noindent 
This research was sponsored by
National Aeronautics and Space Administration, 
National Science Foundation IGERT grant DGE-9987589,
National Research Laboratory Program of KISTEP/MOST (Korea),
Komitet Badan Naukowych Grant 2P03B 032 25 (Poland), 
COFIN MURST, and
Istituto Nazionale di Fisica della Materia (Italy),


\newpage
\noindent
{\bf References}

\begin{description}

\item
Abe, H., and Go, N. 1981.
Noninteracting local-structure model of folding and unfolding transition in
globular proteins. II. Application to two-dimensional lattice proteins.
{\it Biopolymers} {\bf 20:} 1013-1031.     

\item
Abkevich, V.I., Gutin, A.M., and Shakhnovich, E.I. 1994.
Specific nucleus as the transition state for protein folding:
evidence from the lattice model.
{\it Biochemistry} {\bf 33:} 10026-10036. 

\item
Anfinsen, C. 1973.
Principles that govern the folding of protein chains.
{\it Science} {\bf 181:} 223-230.

\item
Baldwin, R.L., and Rose, G.D. 1999a.
Is protein folding hierarchic? I. Local structure and peptide folding.
{\it Trends Biochem. Sci.} {\bf 24:} 26-33.

\item
Baldwin, R.L., and Rose, G.D. 1999b.
Is protein folding hierarchic? II. Folding intermediates and transition
states.
{\it Trends Biochem. Sci.} {\bf 24:} 77-83.

\item
Becker, O.M., and Karplus, M. 1997.
The topology of multidimensional potential energy surfaces: 
theory and application to peptide structure and kinetics.
{\it J. Chem. Phys.} {\bf 106:} 1495-1517.

\item
Bruscolini, P., and Pelizzola, A. 2002.
Exact solution of the Munoz-Eaton model for protein folding.
{\it Phys. Rev. Lett.} {\bf 88:} 258101.

\item
Carmichael Wallace, J.A., editor. 1999.
{\it Protein engineering by semisynthesis}.
CRC Press, New York.

\item
Chan, H.S., and Dill, K.A. 1998.
Protein folding in the 
landscape perspective: chevron plots and non-Arrhenius kinetics.
{\it Proteins} {\bf 30:} 2-33.

\item
Cieplak, M., Henkel, M., Karbowski, J., and Banavar, J.R. 1998.
Master equation approach to protein folding and kinetic traps.
{\it Phys. Rev. Lett.} {\bf 80:} 3654-3657.

\item
Cleland, J.L., and Craik, C.S., editors. 1996.
{\it Protein Engineering: Principles and Practice},
Wiley-Liss, New York.

\item
Clementi, C., Nymeyer, H., and Onuchic, J.N. 2000.
Topological and energetic factors: What determines the structural
details of the transitioon state ensemble and "en-route"
intermediates for protein folding? An investigation for small globular
proteins.
{\it J. Mol. Biol.} {\bf 298:} 937-953.

\item
Dalby, P.A., Oliveberg, M., and Fersht, A.R. 1998.
Folding intermediates of wild-type and mutants of barnase. I. Use
of $\phi$-value analysis and $m$-values to probe the 
cooperative nature of the folding pre-equilibrium.
{\it J. Mol. Biol.} {\bf 276:} 625-646.

\item
Ding, F., Dokholyan, N.V., Buldyrev, S.V., Stanley, H.E., 
and Shakhnovich, E.I. 2002.
Direct molecular dynamics observation of protein folding
transition state ensemble.
{\it Biophys. J.} {\bf 83:} 3525-3532.

\item
Dinner, A., Lazardis, T., and Karplus, M. 1999. Understanding
beta-hairpin formation. Proc. Natl. Acad. Sci. USA 96: 9068-9073.

\item
Dobson, C.M., Sali, A., and Karplus, M. 1998.
Protein folding: a perspective from theory and experiment.
{\it Angew. Chem. Int. Ed.} {\bf 37:} 868-893.

\item
Dokholyan, N.V., Buldyrev, S.V., Stanley, H.E., and
Shakhnovich, E.I. 2000.
Identifying the protein folding nucleus using
molecular dynamics simulations. 
{\it J. Mol. Biol.} {\bf 296:} 1183-1188.

\item
Eyring, H., and Stern, A.E. 1939.
The application of the theory of absolute reaction rates to proteins,
{\it Chem. Rev.} {\bf 24:} 253-270.

\item
Fersht, A.R. 1997.
Nucleation mechanisms in protein folding.
{\it Curr. Opin. Struct. Biol.} {\bf 7:} 3-9.

\item
Fersht, A.R. 1998.
{\it Structure and Mechanism in Protein Science: A Guide
to Enzyme Catalysis and Protein Folding}. New York, Freeman.

\item
Flammini, A., Banavar, J.R., and Maritan, A. 2002.
Energy landscape and native-state structure of proteins - a
simplified model.
{\it Europhys. Lett.} {\bf 58:} 623-629.

\item
Itzhaki. L.S., Otzen, D.E., and  Fersht, A.R. 1995.
The structure of the transition state for folding of chymotrypsin
inhibitor 2 analysed by protein engineering methods: evidence for
a nucleation-condensation mechanism for protein folding.
{\it J. Mol. Biol.} {\bf 254:} 260-288.

\item
Jackson, S.E., and Fersht, A.R. 1991.
Folding of 
chymoptrypsin inhibitor 2. 1. Evidence for a two-state transition.
{\it Biochem.} {\bf 30:} 10428-10435.

\item
Kazmirski, S.L., Wong, K-B., Freund, S.M.V.,
Tan, Y-J., Fersht, A.R., and Daggett, V. 2001.
Protein folding from a highly disordered denatured state: the folding pathway
of chymotrypsin inhibitor 2 at atomic resolution.
{\it Proc. Natl. Acad. Sci.} (USA) {\bf 98:} 4349-4354.

\item
Klimov, D.K., and Thirumalai, D. 2001.
Multiple protein folding nuclei and the transition state ensemble in 
two-state proteins.
{\it Proteins} {\bf 43:} 465-475.

\item
Levy, Y., Jortner, J., and Becker, O.M. 2001.
Solvent effects on the energy landscapes and folding kinetics of
polyalanine.
{\it Proc. Natl. Acad. Sci. (USA)} {\bf 98:} 2188-2193.

\item
Li, A., and Daggett, V. 1994.
Characterization of the transition state of protein
unfolding by use of molecular dynamics: chymotrypsin inhibitor 2.
{\it Proc. Natl. Acad. Sci. (USA)}  {\bf 91}: 10430-10434.

\item
Li, L., Mirny, A.L., and Shakhnovich, E.I. 2000.
KInetics, thermodynamics and evolution of non-native interactions
in a protein folding nucleus.
{\it Nature Struct. Biol.} {\bf 7:} 336-342.

\item
Matouschek, A., Kellis Jr., J.T., Serrano, L., and Fersht, A.R. 1989.
Mapping the transition state and pathway of protein folding by 
protein engineering.
{\it Nature} {\bf 342:} 122-126.

\item
Matouschek, A., Kellis Jr., J.T. Serrano, L.,
Bycroft, M., and  Fersht, A.R.  1990.
Transient folding intermediates characterized
by protein engineering.
{\it Nature} {\bf 346:} 440-445.

\item
Matouschek, A., Serrano, L., and Fersht, A.R. 1992.
The folding of an enzyma IV. Structure of an intermediate in the
refolding of barnase analyzed by a protein engineering procedure,
{\it J. Mol. Biol.} {\bf 224:} 819-835.

\item
Munoz, V., Thompson, P.A., Hofrichter, J., and Eaton, W.A. 1997.
Folding dynamics and mechanism of $\beta$-hairpin formation.
{\it Nature}  {\bf 390:} 196-199.

\item
Munoz, V., Henry, E.R., Hofrichter, J., and Eaton, W.A. 1998.
A statistical mechanical model for $\beta$-hairpin kinetics.
{Proc. Natl. Acad. Sci.} {\bf 95:} 5872-5879.

\item
Munoz, V., and Eaton, W.A. 1999.
A simple model for calculating the kinetics of protein folding
form three-dimensional structures.
{\it Proc. Natl. Acad. Sci. (USA)} {\bf 96:} 11311-11316.

\item
Oliveberg, M., Tan, Y-J., and Fersht, A.R. 1995.
Negative activation enthalpies in the kinetics of protein folding,
{\it Proc. Natl. Acad. Sci. (USA)} {\bf 92:} 8926-8929.

\item
Oliveberg, M. 2001.
Characterisation of the transition states for protein
folding: towards a new level of mechanistic detail in protein
engineering analysis.
{\t Curr. Op. Str. Biol.} {\bf 11:} 94-100.

\item
Onuchic, J.N., Wolynes, P.G., Luthey-Schulten, Z., and Socci, N.D. 1995.
Toward an outline of the topography of a realistic protein-folding
funnel.
{\it Proc. Natl. Acad. Sci. (USA)} {\bf 92:} 3626-3630.

\item
Otzen, D.S., ElMasry, N., Jackson, S.E., and Fersht, A.R. 1994.
The structure of the transition state for the folding/unfolding
of the barley chymotrypsin inhibitor 2 and its implications for
mechanisms of protein folding.
{\it Proc. Natl. Acad. Sci. (USA)} {\bf 91:} 10422-10425.

\item
Oxender, D.L., and Fox, C.F., editors. 1987.
{\it Protein Engineering}, John Wiley \& Sons New York.

\item
Ozkan, S.B.,  Bahar, I.,  and Dill, K.A. 2001.
Transition states and the meaning
of $\phi$-values in protein folding kinetics.
{\it Nature Struct. Biol.} {\bf 8:} 765-769.

\item
Ozkan, S.B., Dill, K.A., and Bahar, I. 2002.
Fast-folding protein kinetics, hidden inetrmediates, and
the sequential stabilization model.
{\it Prot. Sci.} {\bf 11:} 1971-1977.

\item
Paci, E., Clarke, J., Steward, A., Vendruscolo, M., and Karplus, M. 2003.
Self-consistent determination of the transition state for
protein folding: application to a fibronectin type III domain.
{\it Proc. Natl. Acad. Sci. (USA)} {\bf 100:} 394-399.

\item
Pande, V.S., Grosberg, A.Yu., Tanaka, T., and Rokhsar, D.S. 1998.
Pathways for protein folding: is a new view needed.
{\it Curr. Opin. Struct. Biol.} {\bf 8:} 68-79.

\item
Pande, V.S., and Rokhsar, D.S. 1999a.
Folding pathway of a lattice model for proteins.
{\it Proc. Natl. Acad. Sci. (USA)} {\bf 96:} 1273-1278.

\item
Pande, V.S., and Rokhsar, D.S. 1999b.
Molecular dynamics simulations of unfolding and refolding of a
$\beta$-hairpin fragment of protein G.
{\it Proc. Natl. Acad. Sci. (USA)}  {\bf 96:} 9062-9067.

\item
Robson, B., and Garnier, J. 1988.
{\it Introduction to proteins and protein engineering.}
Elsevier Science Ltd. New York.

\item
Shea, J.E., Onuchic, J.N., and Brooks, C.L. 2000.
Energetic frustration and the nature of the transition state in
protein folding.
{\it J. Chem. Phys.} {\bf 113:} 7663-7671.

\item
Ternstroem, T., Mayor, U., Akke, M., and Oliveberg, M. 1999.
>From snapshot to movie: $\phi$ analysis of protein folding transition
states taken one step further.
{\it Proc. Natl. Acad. Sci. (USA)} {\bf 96:} 14854-14859.

\item
Tissot, A.C., Vuilleumier, S., and Fersht, A.R. 1996.
Inportance of two buried salt bridges in the stability and folding
pathway of barnase.
{\it Biochemistry} {\bf 35:} 6786-6794.

\item
Vendruscolo, M., Paci, E., Dobson, C.M., and Karplus, M. 2001.
Three key residues form a critical contact network in a
protein folding transition state.
{\it Nature} {\bf 409:} 641-645.

\item
Wales, D.J., Doye, J.P.K., Miller, M.A., Mortenson, P.N., and Walsh, T.R.
2000.
Energy landscapes: from clusters to biomolecules.
{\it Adv. Chem. Phys.} {\bf 115:} 1-111.


\end{description}

\newpage

\begin{figure}
\includegraphics[width=8cm]{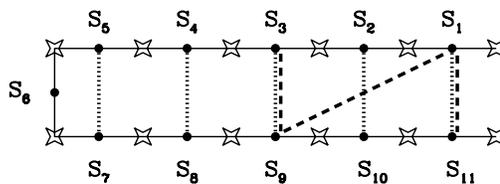}
\caption{ 
The model $\beta$-hairpin system studied in this paper. The stars denote
amino acids. The spins $S_n$ correspond to the peptide bonds between
the successive amino acids. 
In non-native conformations only parts of the native structure
are established. The dotted lines indicate presence
of a hydrogen bond. The dashed lines correspond to hydrophobic
bonds between hydrophobic amino acids.
}
\end{figure}

\begin{figure}
\includegraphics[width=8cm]{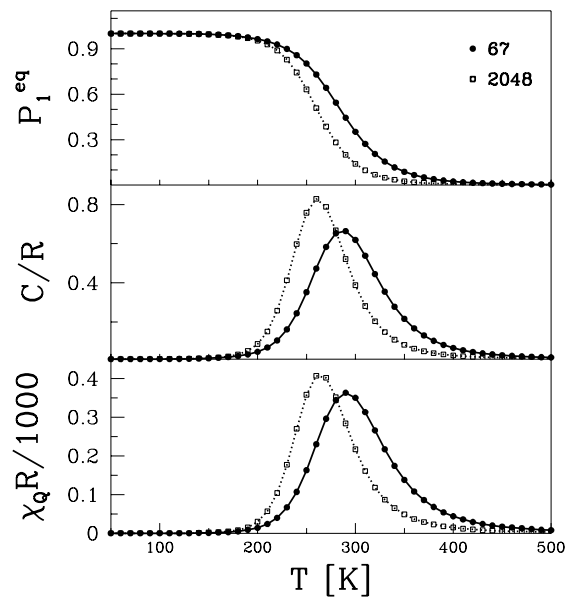}
\caption{ 
The thermodynamic properties of the system. The solid lines
correspond to the single sequence approximation and the dotted lines
to the all-state calculation.
The top panel shows
the equilibrium occupancy of the native state.
The middle panel shows the specific heat, and the bottom panel
the "contact susceptibility", i.e. the fluctuation in the
fraction of the native contacts divided by $RT$
}
\end{figure}

\begin{figure}
\includegraphics[width=8cm]{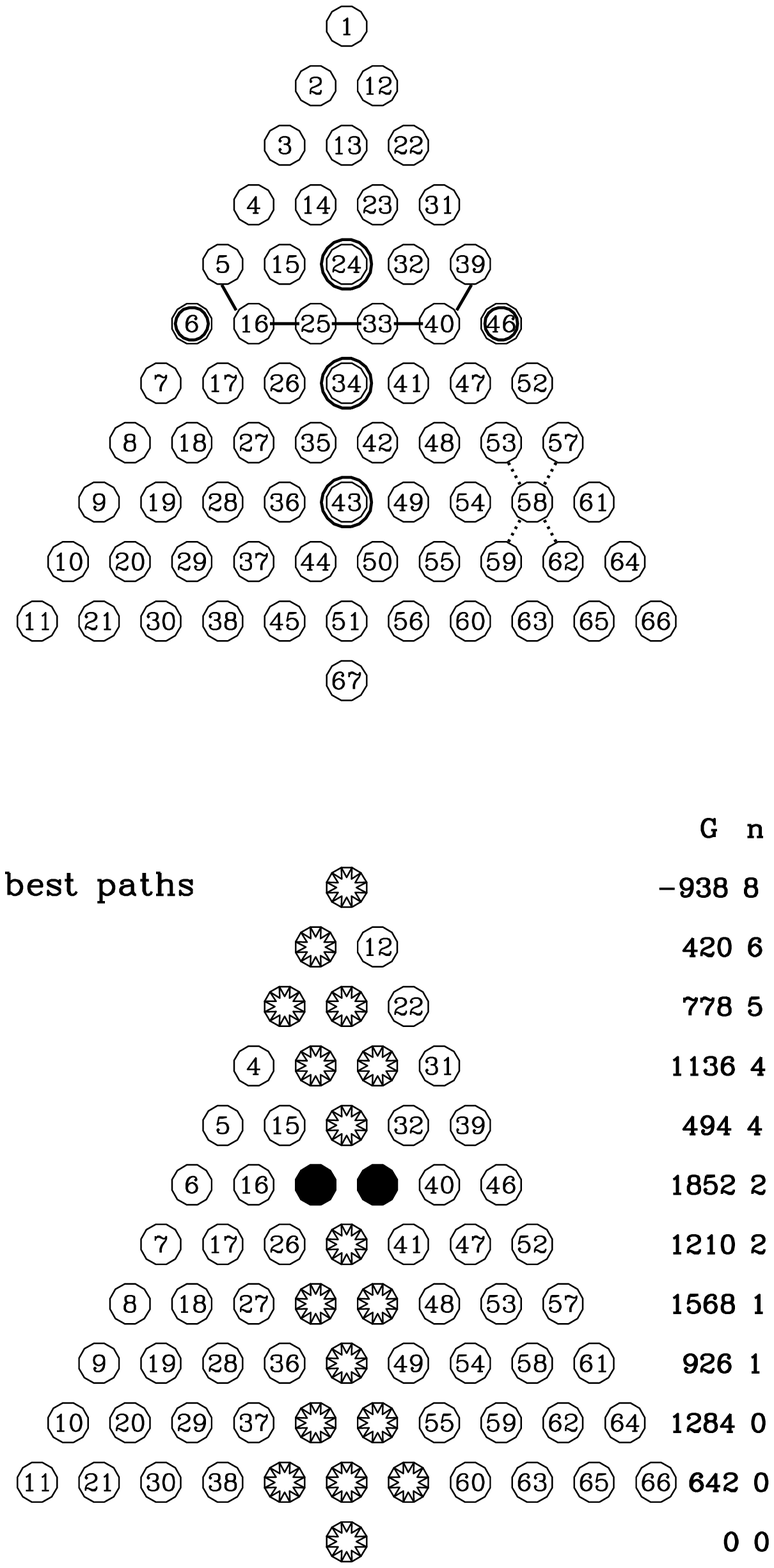}
\caption{ 
A triangular representation of the 67-level system.
The explanations are in the
main text of the paper. The values of the free energies
and of the contact numbers shown on the right of the
bottom panel refer only to the states along the optimal
path and not to all states in each row.
}
\end{figure}

\begin{figure}
\includegraphics[width=8cm]{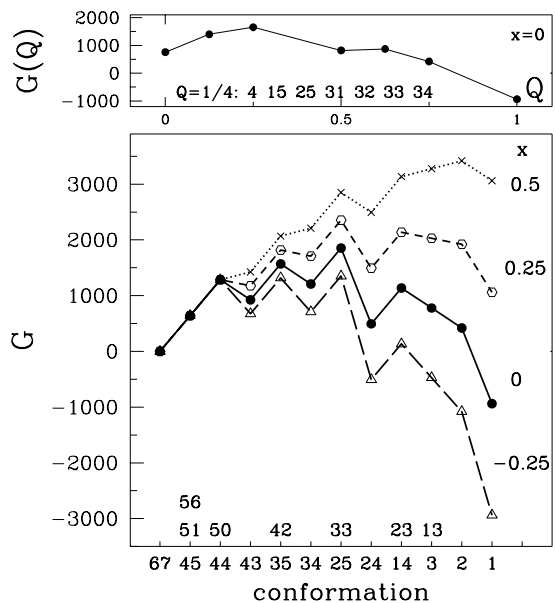}
\caption{ 
The bottom panel shows variations in the free energy
along the optimal paths for four values of the concentration
of the denaturant, $x$. The contact energies are assumed to be
$J(x)=(1-x)J$.  For $x$=0,  the
the free energy landscape is shown in Figure 3.
The reaction coordinate consists of the conformation label(s)
shown at the bottom. The states
25 and 33 are the transition states for the three 
lowest $x$ values shown. For $x$=0.5 it is the almost folded state
2 which becomes the transition state.
The top panel shows the free energy, $G(Q)$ as a function of the
contact number $Q$.
It is obtained by grouping all states into clusters having a given $Q$
and by calculating the average free energy
within each cluster with the normalized Boltzmann factors
as the statistical weights. The maximum of $G(Q)$ occurs at
$Q$=1/4 and corresponds to seven conformations with two contacts each.
}
\end{figure}

\begin{figure}
\includegraphics[width=8cm]{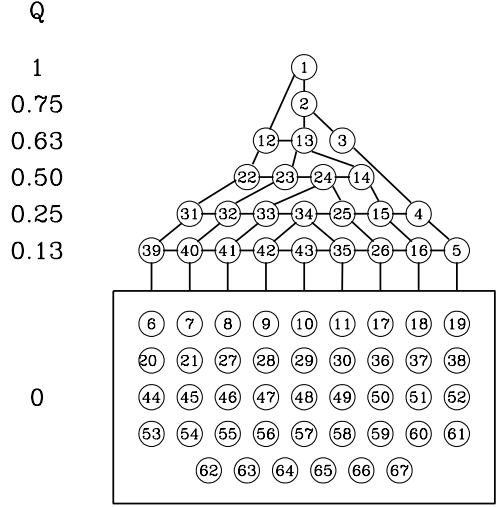}
\caption{ 
The kinetic connectivities in the 67-level system corresponding to 
the scheme in which the states are arranged according to their
$Q$ values. The $Q$ values are indicated on the left-hand side.
The connectivities to and within the states with $Q$=0 are complicated 
and thus not shown.
For example, the kinetic moves from state 58 to states
53, 57, 59 and 62, indicated
by the dotted lines in Figure 3, are all between states with $Q$=0
even though they correspond to a varying magnetization.
Similarly, the moves from state 34 to the two transition states
enhance the magnetization but keep $Q$ at the value of 0.25.
}
\end{figure}


\begin{figure}
\includegraphics[width=8cm]{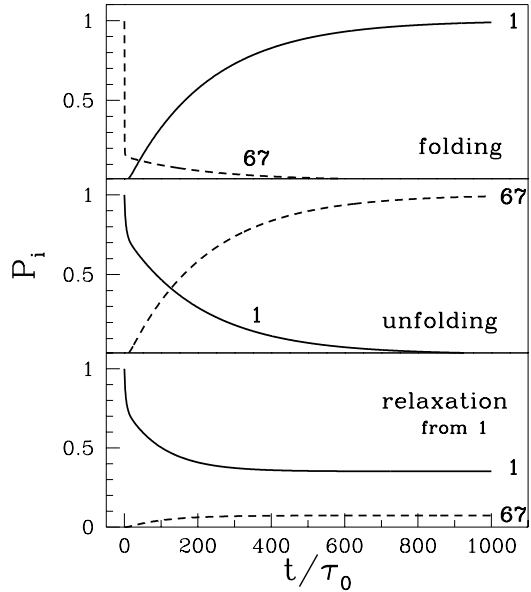}
\caption{ 
Time evolution of the the probability to occupy the native
(solid line) and unfolded (dashed line) states in the 67-level system. 
The initial state of the system is the unfolded state in the top panel
and the folded state in the bottom two panels. }
\end{figure}

\begin{figure}
\includegraphics[width=8cm]{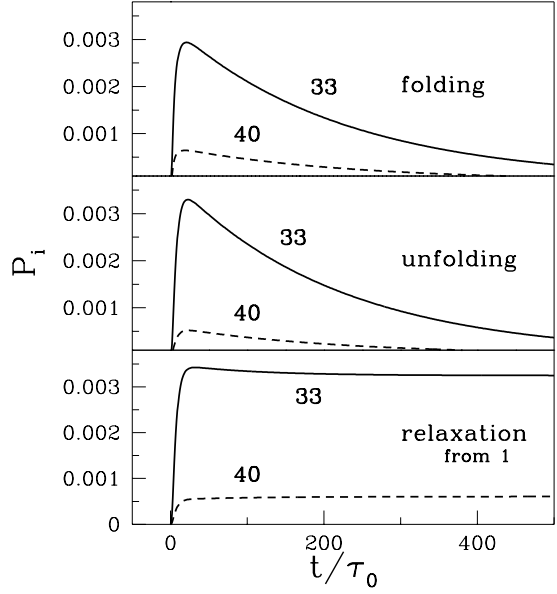}
\caption{ 
Same as in Figure 5 but for the transition state 33 and a nearby
state 40.
}
\end{figure}


\begin{figure}
\includegraphics[width=8cm]{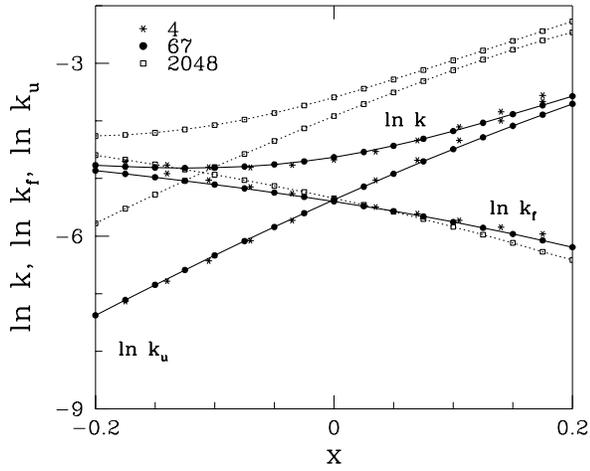}
\caption{ 
Logarithms of the folding, unfolding, and relaxation rates in the
4-level (asterisks), 67-level (circles), and 2048-level (squares)
systems as a function of $x$ in a model in which $J$ is adjusted
linearly by $x$.
The prefactor in the 4-level system was adjusted by a factor of 
4.06 downwards to match the data for the 67-level system. 
}
\end{figure}

\begin{figure}
\includegraphics[width=8cm]{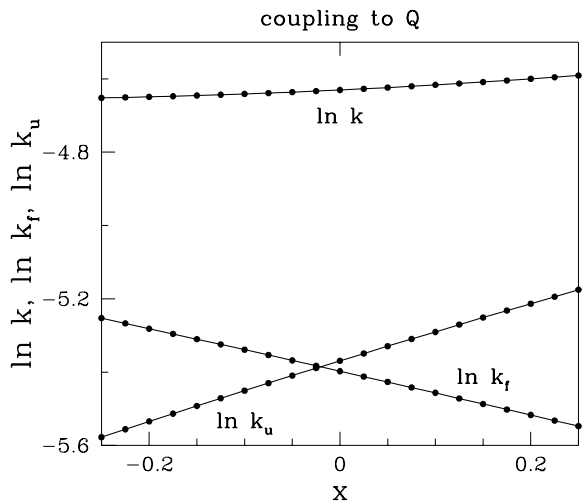}
\caption{ 
Logarithms of the folding, unfolding, and relaxation rates
in the 67-level system as a function of $x$ in a model
in which the free energies of the levels are adjusted
in proportion to $Q$. 
}
\end{figure}

\begin{figure}
\includegraphics[width=8cm]{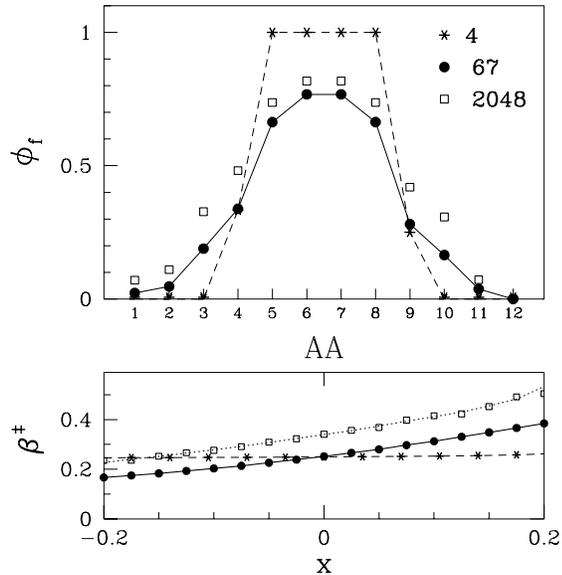}
\caption{ 
The top panel shows $\phi$-values  as obtained in the 4- (asterisks),
67- (circles), and 2048-level (squares) systems. AA stands for the
location of an ``amino acid" where a mutation is implemented.
The bottom panel shows $\beta ^{\ddagger}$ as a function of
denaturant concentration, $x$,
for the three models. The $x$ enters through an adjustment in $J$
due to the denaturant (see also Figure 4). 
}
\end{figure}


\end{document}